\documentclass[12pt,eqsecnum]{article}
\usepackage{graphicx}
\usepackage{amssymb}
\usepackage{amsmath}
\usepackage{ulem}
\usepackage[dvipsnames]{xcolor}
\usepackage{soul}
\usepackage{ascmac}
\usepackage{cite}

\makeatletter

\@addtoreset{equation}{section}
\makeatletter

\newcommand{\qed}{\hbox{\rule[-2pt]{6pt}{6pt}}}
\newcommand{\D}{{\rm d}}

\newtheorem{Prop}{Proposition}

\newtheorem{Coro}{Corollary}

\newcommand{\dalm}{\kern1pt\vbox{\hrule height 0.9pt\hbox{\vrule width
0.9pt\hskip 2.5pt\vbox{\vskip 5.5pt}\hskip 3pt\vrule width 0.3pt}\hrule height
0.3pt}\kern1pt}

\begin{document}

\begin{titlepage}
\vfill
\begin{flushright}
\today
\end{flushright}

\vfill
\begin{center}
\baselineskip=16pt
{\Large\bf
Charged rotating BTZ solution revisited: \\[2mm]
New coordinates and algebraic classifications
}
\vskip 0.3cm
{\large {\sl }}
\vskip 8mm
{\bf Hideki Maeda${}^{a,b}$ and Ji\v{r}\'{\i} Podolsk\'y$^{c}$}

\vskip 5mm
{
${}^a$Department of Electronics and Information Engineering, Hokkai-Gakuen University, Sapporo 062-8605, Japan.\\
${}^b$Max-Planck-Institut f\"ur Gravitationsphysik (Albert-Einstein-Institut), \\Am M\"uhlenberg~1, D-14476 Potsdam, Germany.\\
${}^c$Institute of Theoretical Physics, Charles University, \\
V~Hole\v{s}ovi\v{c}k\'ach 2, 18000 Prague 8, Czechia.\\
\texttt{h-maeda@hgu.jp, jiri.podolsky@mff.cuni.cz}
}
\vspace{6pt}
\end{center}
\vskip 0.2in
\par
\begin{center}
{\bf Abstract}
\end{center}
\begin{quote}
We revisit the charged rotating Ba\~nados-Teitelboim-Zanelli (BTZ) solution in the three-dimensional Einstein-Maxwell-$\Lambda$ system. After the erroneous announcement of its discovery at the end of the original BTZ paper in 1992, the solution was first obtained by Cl\'ement in the paper published in 1996 by coordinate transformations from the charged non-rotating BTZ solution. While Cl\'ement's form of the solution is valid only for ${\Lambda<0}$, we present a new form for a wider range of $\Lambda$ by uniform scaling transformations and a reparametrization. We also introduce new coordinates corresponding to the Doran coordinates in the Kerr spacetime, in which the metric and also its inverse are regular at the Killing horizon, and described by elementary functions. Lastly, we show that (i) the algebraic Cotton type of the spacetime is type III on the Killing horizon and type I away from the horizon, and (ii) the energy-momentum tensor for the Maxwell field is of the Hawking-Ellis type I everywhere.
\vfill
\vskip 2.mm
\end{quote}
\end{titlepage}




\tableofcontents

\newpage

\section{Introduction}

In three spacetime dimensions, the numbers of independent components of the Riemann tensor and the Ricci tensor are the same, so that the Ricci tensor determined by the Einstein equations contains all the local information of spacetime.
This means that three-dimensional gravity is locally trivial and, for example, does not admit gravitational waves encoded in the Weyl tensor.
Therefore, the discovery of the Ba\~nados-Teitelboim-Zanelli (BTZ) solution~\cite{Banados:1992wn} with a negative cosmological constant $\Lambda$ was greeted with great surprise by the community because the spacetime is locally anti-de~Sitter (AdS) but can still describe a rotating black hole globally by proper identifications of spacetime events. Similar to the Kerr-AdS black hole in four dimensions, the rotating BTZ black hole in asymptotically AdS possesses an inner horizon and admits closed time curves (in the region where ${r^2<0}$ holds)~\cite{Banados:1992gq}. For this reason, it has been intensively studied in order to gain insight into the quantum theory of gravity~\cite{Carlip:1995qv}.

The charged generalization of the rotating BTZ solution with a non-trivial Maxwell field has, in fact, a noteworthy complicated history. Its discovery was first mentioned in 1992 at the end of the original BTZ paper~\cite{Banados:1992wn} on the rotating BTZ solution in the vacuum case, but unfortunately the metric and gauge field do not satisfy the field equations in the rotating case.\footnote{See the note added at the end of the arXiv version of the paper~\cite{Banados:1992wn}.} The charged {\it non-rotating} BTZ solution was discovered independently by Peldan in the same year~\cite{Peldan:1992mp}.
After several years, the charged rotating BTZ solution was first obtained by Cl\'ement in the paper published in 1996~\cite{Clement:1995zt} by coordinate transformations from the non-rotating solution. Later, in 2000, the solution was studied in detail by Mart\'{\i}nez, Teitelboim, and Zanelli~\cite{Martinez:1999qi}. It should be noted that the charged rotating solution obtained earlier by Cl\'ement (Eq.~(23) in~\cite{Clement:1993kc}) is also locally identical to the charged non-rotating BTZ solution, but a double Wick rotation is required. (See Appendix~\ref{app:Clement}.)
On the other hand, the charged rotating solution obtained by Kamata and Koikawa in~\cite{Kamata:1995zu}, slightly earlier than~\cite{Clement:1995zt}, belongs to a different class from the charged BTZ solution because it gives $F_{\mu\nu}F^{\mu\nu}=0$. (See Sec.~11.6 in~\cite{Garcia-Diaz:2017cpv}.) Unfortunately, a non-negligible number of papers have been published wrongly using the incorrect metric of the charged rotating BTZ solution presented in~\cite{Banados:1992wn} for the analysis in the Einstein-Maxwell-$\Lambda$ system, and such papers appear even at present~\cite{wrong1,wrong2,Cadoni:2007ck,Larranaga:2008vt,wrong3,wrong4,wrong5,Hussain:2011zzb,Kawamoto:2017ufe,wrong6,Sharif:2019nmn,wrong8,wrong9,Hamil:2022bpd}.

Despite this situation, the understanding of the solution space of the Einstein equations in three dimensions has progressed rapidly in recent years. In particular, the Einstein equations with $\Lambda$ have been solved with only few natural assumptions imposed, and the local structure of the solutions has been classified not only in the vacuum case but also in the presence of a null dust fluid or a gyratonic matter~\cite{Podolsky:2018zha}. In the Einstein-Maxwell-$\Lambda$ system, stationary and circularly symmetric solutions have been classified by Garc\'\i{}a-D\'\i{}az~\cite{Garcia-Diaz:2013aaa,Garcia:2009vsk}, which (locally) include the charged rotating BTZ solution. Recently, the field equations in this system have been fully solved without imposing any spacetime symmetry~\cite{Podolsky:2021gsa}. Exact solutions to the Einstein equations in three dimensions until 2017 have been summarized in~\cite{Garcia-Diaz:2017cpv}.

It should be emphasized here that the coordinate system of the charged rotating BTZ solution derived in~\cite{Martinez:1999qi} does not cover the Killing horizons corresponding to the event horizon and the inner horizon of a black hole, and therefore does not directly allow for a correct analysis of some geometrical and physical properties on the horizon. This fact can be most easily exhibited in the calculation of surface gravity $\kappa$ of the famous four-dimensional Schwarzschild black hole using the diagonal coordinates
\begin{align}
\label{Schw}
&\D s^2=-\biggl(1-\frac{2GM}{r}\biggl)\,\D t^2+\biggl(1-\frac{2GM}{r}\biggl)^{-1}\D r^2+r^2(\D\theta^2+\sin^2\theta\,\D \phi^2).
\end{align}
The surface gravity $\kappa$ on the Killing horizon at ${r=2GM}$ associated with the Killing vector $\xi^\mu(\partial/\partial x^\mu)=\partial/\partial t$ is defined by ${\xi^\nu\nabla_\nu\, \xi^\mu|_{r=2GM}=\kappa\,\xi^\mu|_{r=2GM}}$.
Actually, the left-hand side is zero for all $\mu$ in the coordinates (\ref{Schw}), so that one arrives at a wrong answer ${\kappa=0}$. 
Of course, the reason for this error is that the coordinates (\ref{Schw}) do not cover the horizon.
In the maximally extended Schwarzschild spacetime, the black-hole type future Killing horizon corresponds to $(t,r)\to (+\infty,2GM)$, while the white-hole type past Killing horizon corresponds to $(t,r)\to (-\infty,2GM)$.
In fact, $(t,r)\to (t_0,2GM)$ with a finite constant $t_0$ corresponds to a bifurcation two-sphere where the Killing vector generating staticity becomes a zero vector.
With the advanced time ${v:=t+\int (1-2GM/r)^{-1}\D r}$ and $\xi^\mu(\partial/\partial x^\mu)=\partial/\partial v$ instead of $\partial/\partial t$, one obtains a correct answer ${\kappa=1/(4GM)}$ for the black-hole horizon, while one obtains ${\kappa=-1/(4GM)}$ for the white-hole horizon with the retarded time ${u:=t-\int (1-2GM/r)^{-1}\D r}$ and ${\xi^\mu(\partial/\partial x^\mu)=\partial/\partial u}$.

The necessity of a regular coordinate system covering the Killing horizon was recently emphasized in the Hawking-Ellis classification of the energy-momentum tensor in static spacetimes with symmetry~\cite{Maeda:2021ukk} and also in the Petrov classification of spacetime~\cite{Pravda:2005uv,Tanatarov:2012gf}.
For these reasons, it is undoubtedly important to find a coordinate system that covers a horizon in exact solutions.
However, it is not easy even for a stationary and axisymmetrc spacetime to find coordinates in which the metric and its inverse are not only finite on the horizon but also described by elementary functions. In the case of the four-dimensional Kerr spacetime, the Doran coordinates are an example of such coordinates~\cite{Doran:1999gb}, which reduce to the Painlev\'{e}-Gullstrand coordinates for the Schwarzschild spacetime in the non-rotating limit.
Until now, the Doran-type horizon-penetrating coordinates have also been obtained for the Kerr-Newman spacetime~\cite{Lin:2012ziw,Hobson:2022cso}, uncharged rotating BTZ spacetime~\cite{Zaslavskii:2018lbb}, and five-dimensional Myers-Perry spacetime~\cite{Finch:2013gha}. We note that, in~\cite{Lin:2012ziw}, although the authors presented coordinates with $\Lambda$, namely in the Kerr-Newman-(A)dS spacetime, they are not horizon-penetrating with $\Lambda$ as the inverse metric diverges at the Killing horizon. In~\cite{Zaslavskii:2018lbb}, the author studied horizon-penetrating coordinates in the generic stationary and axisymmetric spacetime in four dimensions. But again, the finiteness of the inverse metric was not taken into account.

In the present paper, we will revisit the charged rotating BTZ solution and expose its several new aspects.
The organization of the article is as follows. 
After introducing the field equations and reviewing axisymmetric spacetimes in three dimensions in Sec.~\ref{sec:review}, we will describe our main result in Sec.~\ref{sec:main}.
We will first present a new form of the charged rotating BTZ solution that is valid for a wider range of $\Lambda$, and then derive horizon-penetrating coordinates in a much more general spacetime. 
Then, we will perform the recently established Cotton algebraic classification of spacetime~\cite{Podolsky:2023qiu,Papajcik:2023zen} and the Hawking-Ellis classification~\cite{Hawking:1973uf,Maeda:2018hqu} of the energy-momentum tensor for the Maxwell field.
Concluding remarks are given in the final section. 
Appendix~\ref{app:Clement} shows that Cl\'ement's solution given by Eq.~(23) in~\cite{Clement:1993kc} is locally identical to the charged non-rotating BTZ solution. 
Appendix~\ref{app:derivation} explains the derivation of the Doran-type new coordinates. 
Throughout this article, the signature of the Minkowski spacetime is $(-,+,+)$. We adopt the units such that ${c=1}$, and the conventions of curvature tensors such as ${[\nabla _\rho ,\nabla_\sigma]V^\mu ={{\cal R}^\mu }_{\nu\rho\sigma}V^\nu}$ and ${{\cal R}_{\mu \nu }={{\cal R}^\rho }_{\mu \rho \nu }}$. We use ${\kappa:=8\pi G}$ instead of the gravitational constant $G$.

\section{Einstein-Maxwell-$\Lambda$ system in three dimensions}
\label{sec:review}

\subsection{Field equations}

In the present paper, we study the charged rotating BTZ solution in the three-dimensional Einstein-Maxwell-$\Lambda$ system.
The action for the spacetime metric $g_{\mu\nu}$ and the $U(1)$ gauge field $A_\mu$ is given by
\begin{align}
\label{action}
S[g_{\mu\nu}, A_{\mu}]=&\int \D^3x\,\sqrt{-g}\,\biggl(\frac{1}{2\kappa}({\cal R}-2\Lambda)-\frac{1}{4}F_{\mu\nu}F^{\mu\nu}\biggl)+S_\Sigma,
\end{align}
where $F_{\mu\nu}:=\nabla_\mu A_\nu-\nabla_\nu A_\mu$ is the Faraday tensor and $S_\Sigma$ is the boundary term.
Variation of the action gives the field equations
\begin{align}
G_{\mu\nu}+\Lambda g_{\mu\nu}&=\kappa \,T_{\mu\nu}, \label{EFE}\\
\nabla_\nu F^{\mu\nu}&=0, \label{em-kg}
\end{align}
where $G_{\mu\nu}$ is the Einstein tensor and the energy-momentum tensor $T_{\mu\nu}$ for the Maxwell field is given by
\begin{align}
T_{\mu\nu}=&F_{\mu\rho}F_\nu^{~\rho}-\frac 14 g_{\mu\nu}F_{\rho\sigma}F^{\rho\sigma}. \label{Tab-Max}
\end{align}

We note that the system (\ref{action}) is equivalent to the Einstein-$\Lambda$ system with a massless scalar field $\phi$.
In three dimensions, the dual Maxwell one-form is defined by 
\begin{align}
{}^*F_{\mu}:=\frac12\varepsilon_{\mu\nu\rho}F^{\nu\rho}~~\left(\Leftrightarrow~~{}^*F_{\mu}\varepsilon^{\mu\alpha\beta}=-F^{\alpha\beta}\right),\label{dualF}
\end{align}
where the totally anti-symmetric volume three-form $\varepsilon_{\mu\nu\rho}$ is defined by 
\begin{align}
\varepsilon_{\mu\nu\rho}:=\sqrt{-g}\epsilon_{\mu\nu\rho}~~\left(\Leftrightarrow~~\varepsilon^{\mu\nu\rho}=-\epsilon^{\mu\nu\rho}/\sqrt{-g}\right)
\end{align}
with the Levi-Civita symbol $\epsilon_{\mu\nu\rho}$ satisfying $\epsilon_{012}=1$ and $\epsilon^{012}=1$.
Using the Maxwell equations and $\varepsilon^{\mu\alpha\beta}\varepsilon_{\mu\nu\rho}=-\epsilon^{\mu\alpha\beta}\epsilon_{\mu\nu\rho}=-(\delta^\alpha_{~\nu}\delta^\beta_{~\rho}-\delta^\alpha_{~\rho}\delta^\beta_{~\nu})$, we obtain
\begin{align}
\varepsilon^{\alpha\sigma\mu}\nabla_\sigma {}^*F_\mu=-\nabla_\sigma F^{\alpha\sigma}=0,
\end{align}
which shows that there exists a potential scalar ${\phi}$ satisfying ${\nabla_\mu \phi= {}^*F_{\mu}}$ by the Poincar{\'e} lemma.
This potential ${\phi}$ is identified as a massless scalar field.
In fact, the equivalence between the energy-momentum tensors of a Maxwell field and a massless scalar field is shown as
\begin{align}
T_{\mu\nu}=F_{\mu\rho}F_\nu^{~\rho}-\frac 14 g_{\mu\nu}F_{\rho\sigma}F^{\rho\sigma}=(\nabla_{\mu}\phi)(\nabla_{\nu}\phi)-\frac 12g_{\mu\nu}(\nabla\phi)^2, \label{Tab-dual}
\end{align}
where ${(\nabla\phi)^2:=(\nabla_{\alpha}\phi)(\nabla^{\alpha}\phi)}$ and ${F_{\mu\nu}F^{\mu\nu}=-2(\nabla\phi)^2}$.
Also, the equation of motion for a scalar field ${\nabla_\mu\nabla^\mu\phi=0}$ is satisfied, shown as
\begin{align}
&\nabla_\mu\nabla^\mu\phi=\frac12\varepsilon_{\mu\nu\rho}\nabla^\mu F^{\nu\rho}=\frac12\varepsilon_{\mu\nu\rho}\nabla^{[\mu}F^{\nu\rho]}=0.
\end{align}

It should be noted that there is a no-go theorem for three-dimensional black holes as solutions to the Einstein equations (\ref{EFE}) without specifying a matter field.
Ida's theorem asserts the absence of an apparent horizon for $\Lambda>0$ under the dominant energy condition (DEC) for the matter field $T_{\mu\nu}$~\cite{Ida:2000jh}.
Here an apparent horizon is the outer boundary of outer trapped regions and the theorem asserts the absence of the event horizon in the stationary case.
In his paper~\cite{Ida:2000jh}, Ida also commented that the same holds for $\Lambda=0$ with a Maxwell field.
Therefore, a black hole is possible only for $\Lambda<0$ in the present system.

\subsection{Axisymmetric solutions}
\subsubsection{KGBD quasi-local mass and angular momentum}

The most general metric in a three-dimensional axisymmetric spacetime without assuming stationarity can be generally written as
\begin{align}
&\D s^2=g_{\mu\nu}\D x^\mu \D x^\nu=h_{ij}(y)\D y^i\D y^j+R(y)^2(\D \theta+a_i(y)\D y^i)^2,\label{general-axial}
\end{align}
where $i,j=0,1$ and $\psi^\mu(\partial/\partial x^\mu)=\partial/\partial \theta$ is a Killing vector generating axisymmetry.
Recently, a quasi-local mass $m$ and a quasi-local angular momentum $j$ for axisymmetric spacetimes in three dimensions have been defined by Gundlach, Bourg, and Davey~\cite{Gundlach:2021six} and independently by Kinoshita~\cite{Kinoshita:2021qsv} in a covariant manner without using the coordinate system (\ref{general-axial}).
Their definitions differ only by constant factors, and we follow Kinoshita's definitions because the quasi-local mass in his definition reduces to the three-dimensional Misner-Sharp quasi-local mass if the Killing vector $\psi^\mu$ is hypersurface-orthogonal, namely, it satisfies $\psi_{[\mu}\nabla_\nu\psi_{\rho]}=0$.
Let us now briefly summarize the results in~\cite{Gundlach:2021six,Kinoshita:2021qsv}.

A three-dimensional axisymmetric spacetime is defined by the existence of a spacelike Killing vector $\psi^\mu$ with a closed orbit. 
The squared norm of $\psi^\mu$ defines the areal radius as
\begin{align}
R:=\sqrt{\psi_\mu\psi^\mu}.
\end{align}
The metric in the reduced two-dimensional spacetime of orbits of $\psi^\mu$ is given by
\begin{align}
h_{\mu\nu}:=g_{\mu\nu}-R^{-2}\psi_\mu\psi_\nu,
\end{align}
which satisfies $h_{\mu\nu}\psi^\nu=0$.
The volume two-form in the reduced spacetime is defined by 
\begin{align}
{\bar\varepsilon}_{\mu\nu}:=R^{-1}\varepsilon_{\mu\nu\rho}\psi^\rho,
\end{align}
which satisfies ${\bar\varepsilon}_{\mu\rho}{\bar\varepsilon}^{\mu\sigma}=-h_\rho^{~\sigma}$, where $\varepsilon_{\mu\rho\sigma}$ is totally anti-symmetric volume three-form.
(See Sec.~2.8 in the textbook~\cite{Carroll:2004st} for the properties of $\varepsilon_{\mu\rho\sigma}$.)

In~\cite{Kinoshita:2021qsv}, Kinoshita defined $m$ and $j$ as
\begin{align}
&m:=\frac{\pi}{\kappa}(-\Lambda\,\psi_\mu \psi^\mu+K_\mu K^\mu),\label{def-QLM2}\\
&j:=\frac{1}{\kappa}\varepsilon^{\mu\rho\sigma}\psi_\mu\nabla_\rho\psi_\sigma\,\biggl(=-\frac{2\pi}{\kappa}\psi_\mu K^\mu\biggl),\label{def-QLJ2}
\end{align}
where $K^\mu$ is the generalized Kodama vector defined by 
\begin{align}
&K^\mu:=-\frac12\varepsilon^{\mu\rho\sigma}\nabla_\rho \psi_\sigma\,\biggl(=-{\bar\varepsilon}^{\mu\nu}\nabla_\nu R-\frac{\kappa j}{2\pi R^2}\psi^\mu\biggl).\label{def-K2}
\end{align}
The vector $K^\mu$ shares the same properties as the Kodama vector in $n(\ge 3)$ dimensions~\cite{Kodama:1979vn,Maeda:2007uu}, namely, (i) $\nabla_\mu K^\mu=0$, (ii) $K^\mu\nabla_\mu R=0$, and (iii) $G^{\mu\nu}\nabla_\mu K_\nu=0$.
If $\psi^\mu$ is hypersurface-orthogonal, $j=0$ holds and then $m$ and $K^\mu$ reduce to the Misner-Sharp quasi-local mass~\cite{Misner:1964je,Maeda:2006pm} and the Kodama vector~\cite{Kodama:1979vn,Maeda:2007uu} in three dimensions ($n=3$), respectively.
The quantities $m$ and $j$ satisfy the relation
\begin{align}
8Gm=&-\Lambda R^2+\frac{(4G j)^2}{R^2}-(\nabla_\mu R)(\nabla^\mu R).\label{rel-mj2}
\end{align}
We will refer to $m$ and $j$ as the Kinoshita-Gundlach-Bourg-Davey (KGBD) quasi-local mass and the KGBD quasi-local angular momentum, respectively.

Similar to the Misner-Sharp mass~\cite{Burnett:1991nh,Burnett:1993bn,Hayward:1994bu,Maeda:2007uu}, the KGBD mass $m$ and angular momentum $j$ satisfy \begin{align}
\nabla_\mu\, m=-2\pi R\,{\bar\varepsilon}_{\mu\rho}\,{\cal J}_{(K)}^\rho~~&\Leftrightarrow~~{\bar\varepsilon}^{\sigma\mu}\nabla_\mu\, m=-2\pi R\,h^{\sigma}_{~\rho}\,{\cal J}_{(K)}^\rho,\label{diffm}\\
\nabla_\mu\, j=-2\pi R\,{\bar\varepsilon}_{\mu\rho}\,{\cal J}_{(\psi)}^\rho~~&\Leftrightarrow~~{\bar\varepsilon}^{\sigma\mu}\nabla_\mu\, j=-2\pi R\,h^{\sigma}_{~\rho}\,{\cal J}_{(\psi)}^\rho,\label{diffj}
\end{align}
where
\begin{align}
{\cal J}_{(K)}^\rho:=-T^{\rho\sigma}K_\sigma,\qquad {\cal J}_{(\psi)}^\rho:=T^{\rho\sigma}\psi_\sigma.
\end{align}
Since ${\cal J}_{(K)}^\mu$ (${\cal J}_{(\psi)}^\mu$) is a divergence-free current vector associated with $K^\mu$ ($\psi^\mu$), the quantity $m$ ($j$) is a locally conserved charge associated with ${\cal J}_{(K)}^\mu$ (${\cal J}_{(\psi)}^\mu$). 
In addition, $m$ and $j$ are constant in vacuum ($T_{\mu\nu}=0$).
As a consequence, with those constants $m$ and $j$, the rotating BTZ vacuum solution~\cite{Banados:1992wn} can be written as
\begin{align}
\label{rotatingBTZ-mj}
\begin{aligned}
&\D s^2=-f\,\D t\,^2+f^{-1}\D r^2+r^2\biggl(\D\theta-\frac{4Gj}{r^2}\D t\biggl)^2,\\
&f(r)=-\Lambda r^2-8Gm+\frac{(4Gj)^2}{r^2}.
\end{aligned}
\end{align}
In Sec.~11.11 of the textbook~\cite{Garcia-Diaz:2017cpv}, Garc\'\i{}a-D\'\i{}az computed the Brown-York quasi-local quantities such as energy, mass, and momentum~\cite{Brown:1992br} for the charged rotating BTZ solution along the methods in~\cite{Brown:1994gs}.
However, unlike the KGBD mass and angular momentum, they are not constant even for the rotating BTZ vacuum solution (\ref{rotatingBTZ-mj}).

In fact, although the Brown-York (quasi-local) mass converges to the ADM (global) mass at spacelike infinity in an asymptotically flat spacetime, it gives an $r$-dependent profile for the Schwarzschild-Tangherlini solution and coincides with the ADM mass only asymptotically $r\to \infty$. (See~\cite{Chakraborty:2015kva}, for example.)
In contrast, the Misner-Sharp (quasi-local) mass was originally defined for spherically symmetric spacetimes in four dimensions without $\Lambda$~\cite{Misner:1964je} and its generalization has been defined for spacetimes with more general symmetries in arbitrary $n(\ge 3)$ dimensions with $\Lambda$~\cite{Maeda:2006pm,Maeda:2007uu}.
Similar to the Brown-York mass, it converges to the ADM mass at spacelike infinity in an asymptotically flat spacetime (for $\Lambda=0$).
However, different from the Brown-York mass, the Misner-Sharp mass is {\it constant} for the Schwarzschild-Tangherlini solution which coincides with the ADM mass.
In addition, it has monotonicity and positivity properties under the dominant energy condition~\cite{Hayward:1994bu,Maeda:2007uu}.
Hence, as they are natural generalizations of the three-dimensional Misner-Sharp mass, we prefer the KGBD mass and angular momentum in the present study rather than the Brown-York mass. (See~\cite{Szabados:2004xxa} for a review of quasi-local quantities in general relativity.)

Depending on the parameters $m$ and $j$, the rotating BTZ vacuum spacetime (\ref{rotatingBTZ-mj}) admits Killing horizons located at ${r=r_{\rm h}}$ determined by $f(r_{\rm h})=0$, namely $r_{\rm h}=r_\pm$, where
\begin{align}
& r_\pm^2= \frac{4Gm}{(-\Lambda)}\, \bigg( 1 \pm \sqrt{ 1 + \Lambda \frac{j^2}{m^2}}\,\, \bigg).
\end{align}
The {\it extremal} rotating BTZ vacuum solution is realized for
\begin{align}
|m|=\sqrt{-\Lambda}\,|j|. \label{extreme-mj}
\end{align}
Then, the radius of the degenerate (extreme) horizon $r=r_{\rm ex}$ is given by 
\begin{align}
r_{\rm ex}:= \sqrt{\frac{4Gm}{(-\Lambda)}}, \label{rex-mj}
\end{align}
which satisfies ${f(r_{\rm ex})=f'(r_{\rm ex})=0}$, where a prime denotes differentiation with respect to ${r}$.

\subsubsection{Two coordinate systems in the stationary case}

The most general metric for stationary and axisymmetric spacetime in three dimensions may be written as
\begin{align}
\label{chargedBTZ-0}
\D s^2=-\frac{r^2}{R^2}\,f\,\D t^2+f^{-1}\D r^2+R^2\biggl(\D\theta+\frac{h}{R^2}\,\D t\biggl)^2
\end{align}
in the suitable coordinates $(t,r,\theta)$, where $f=f(r)$, $R=R(r)$, and $h=h(r)$.
The metric (\ref{chargedBTZ-0}) gives ${\sqrt{-\det g}=r}$, and the following non-zero components of the inverse metric
\begin{align}
g^{tt}=-\frac{R^2}{r^2f},\qquad g^{t\theta}=\frac{h}{r^2f},\qquad g^{rr}=f,\qquad g^{\theta\theta}=\frac{r^2f-h^2}{r^2R^2f}.
\end{align}
In this spacetime, ${g_{tt}(r_{\rm erg})=0}$ determines the radius of an ergocircle ${r=r_{\rm erg}}$, while ${f(r_{\rm h})=0}$ determines locations of Killing horizons $r=r_{\rm h}$ associated with a Killing vector $\xi^\mu=(1,0,-h(r_{\rm h})/R^2(r_{\rm h}))$.

For the metric (\ref{chargedBTZ-0}) with a Killing vector $\psi^\mu=(0,0,1)$ generating axisymmetry, the generalized Kodama vector (\ref{def-K2}) is given by 
\begin{align}
&K^\mu\frac{\partial}{\partial x^\mu}=\frac{RR'}{r}\frac{\partial}{\partial t}-\frac{h'}{2r}\frac{\partial}{\partial \theta},\label{g-Kodama-general}
\end{align}
and the dual one-form is 
\begin{align}
&K_\mu\D x^\mu=-\frac{2r^2fR' + hh'R-2h^2R'}{2rR}\D t-\frac{R(h'R-2hR')}{2r}\D\theta,
\end{align}
with which the KGBD mass (\ref{def-QLM2}) and the KGBD angular momentum (\ref{def-QLJ2}) are computed to give 
\begin{align}
m=&\frac{ R^2}{8G}\biggl[\frac{h^2}{4r^2}\biggl(2\frac{{R'}}{R}-\frac{h'}{h}\biggl)^2-f\frac{{R'}^2}{R^2} -\Lambda\biggl],\label{m-rep}\\
j=&-\frac{hR^2}{8G r}\biggl(2\frac{R'}{R}-\frac{h'}{h}\biggl),\label{j-rep}
\end{align}
where we have used $\kappa=8\pi G$.
Equation~(\ref{rel-mj2}) gives the relation
\begin{align}
8Gm=&-\Lambda R^2+\frac{(4G j)^2}{R^2}-f{R'}^2.
\end{align}

Let us notice that one cannot directly study geometry on the Killing horizon ${r=r_{\rm h}}$ in the coordinate system (\ref{chargedBTZ-0}) because it is {\it singular} there as the metric diverges.
Therefore, in order to perform a correct analysis on the horizon, we need to have a coordinate system in which the metric and its inverse are both regular at ${r=r_{\rm h}}$.
In terms of a new coordinate $x$ defined by
\begin{align}
&x:=\int \frac{r}{R(r)}\,\D r,\label{x-def}
\end{align}
the metric (\ref{chargedBTZ-0}) is written as
\begin{align}
&\D s^2=-H\,\D t^2+H^{-1}\D x^2+{\bar R}^2\biggl(\D\theta+\frac{{\bar h}}{{\bar R}^2}\D t\biggl)^2,
\end{align}
where
\begin{align}
&H(x):=\frac{r(x)^2}{R(r(x))^2}f(r(x)),\qquad {\bar R}(x):=R(r(x)),\qquad {\bar h}(x)\,:=\,h(r(x)).
\end{align}
Then, introducing new coordinates $v$ and $\phi$ defined by
\begin{align}
\D v:=\D t+H^{-1}\D x,\qquad \D\phi:=\D\theta-\frac{{\bar h}}{H{\bar R}^2}\,\D x,
\end{align}
we write the metric as\footnote{Actually, this is the Robinson-Trautman form of the metric in which the coordinate $x$ is an affine parameter along a geometrically privileged null expanding (shear-free and twist-free) geodesic congruence.
This was shown in~\cite{Podolsky:2018zha,Podolsky:2021gsa}, where the retarded coordinate $u$ was employed instead of the advanced coordinate $v$, $x$ was denoted as $r$, and $\phi$ was denoted as~$x$.}
\begin{align}
\D s^2=-H\,\D v^2+2\D v\D x+{\bar R}^2\biggl(\D\phi+\frac{{\bar h}}{{\bar R}^2}\,\D v\biggl)^2.\label{null-coord}
\end{align}
The determinant of this metric is $\det g=-{\bar R}^2$, while non-zero components of the inverse metric are given by
\begin{align}
&g^{vx}=1,\qquad g^{xx}=H,\qquad g^{x\phi}=-\frac{{\bar h}}{{\bar R}^2},\qquad g^{\phi\phi}=\frac{1}{{\bar R}^2}.
\end{align}
As a result, the coordinate system (\ref{null-coord}) is regular at the Killing horizon $x=x_{\rm h}$ defined by $H(x_{\rm h})=0$.

Unfortunately, the metric functions in Eq.~(\ref{null-coord}) cannot be written explicitly in terms of elementary functions for the charged rotating BTZ solution due to the complicated integral (\ref{x-def}) with $R(r)$ given by Eq.~(\ref{Clement-R}) or (\ref{R-def}) below.
For this reason, the coordinate system (\ref{null-coord}) is not so useful for studying some specific aspects of the Killing horizons of the solution.

\section{Charged rotating BTZ solution}
\label{sec:main}

\subsection{Conventional form for $\Lambda<0$}

In~\cite{Clement:1995zt}, Cl\'ement obtained the charged rotating BTZ solution in the system (\ref{action}) with ${\Lambda<0}$ in the coordinates (\ref{chargedBTZ-0}), namely,
\begin{align}
\label{chargedBTZ}
\D s^2=-\frac{r^2}{R^2}\,f\,\D t^2+f^{-1}\D r^2+R^2\biggl(\D\theta+\frac{h}{R^2}\,\D t\biggl)^2.
\end{align}
After suitable reparametrizations, the gauge field $A_\mu(r)$ and the metric functions $R(r)$, $f(r)$, and $h(r)$ in his solution are given by
\begin{align}
&A_\mu\D x^\mu=-\frac{Q}{\sqrt{1-\omega^2}}\ln r\left(\D t-\frac{\omega}{\sqrt{-\Lambda}}\,\D\theta\right),\label{Clement-A}\\[1mm]
&R(r)=\sqrt{r^2+\frac{\omega^2}{(-\Lambda)(1-\omega^2)}\big(M+\kappa Q^2\ln r\big)},\label{Clement-R}\\
&f(r)=-\Lambda r^2-M-\kappa Q^2\ln r,\label{Clement-f}\\
&h(r)=-\frac{\omega}{\sqrt{-\Lambda}\,(1-\omega^2)}\big(M+\kappa Q^2\ln r\big),\label{Clement-h}
\end{align}
which gives 
\begin{align}
F_{\mu\nu}F^{\mu\nu}=-\frac{2Q^2}{r^2}.\label{Clement-F2}
\end{align}
This solution is parametrized by three constants, namely $\omega\,(\ne \pm 1)$, $M$, and $Q$, and requires $\Lambda<0$ in order for the metric to be real.
We note that, although the gauge field $A_\mu$ and the Faraday tensor $F_{\mu\nu}$ become pure imaginary if a condition ${-1<\omega<1}$ is not satisfied, the energy-momentum tensor $T_{\mu\nu}$ remains real even in such a case. 
The metric shows $\lim_{r\to\infty}R^{\mu\nu}_{~~\rho\sigma}=\Lambda(\delta^\mu_{~\rho}\delta^\nu_{~\sigma}-\delta^\mu_{~\sigma}\delta^\nu_{~\rho})$, so that the spacetime is asymptotically locally AdS as $r\to \infty$.
We refer to the solution given by Eqs.~(\ref{chargedBTZ})--(\ref{Clement-h}) as the charged rotating BTZ solution in the {\it Cl\'ement form}. In~\cite{Martinez:1999qi}, the solution (\ref{chargedBTZ})--(\ref{Clement-h}) was studied in detail adopting the units $c=-\Lambda=1$.

The Ricci scalar is computed to give
\begin{align}
{\cal R}=6\Lambda+\frac{\kappa Q^2}{r^2},
\end{align}
which shows that $r=0$ is a scalar polynomial curvature singularity for $Q\ne 0$.
As a consequence, the domain of $r$ for the charged rotating BTZ solution is determined by the constraint ${R^2(r)>0}$.
The charged solution (${Q\ne 0}$) describes a black hole only for ${\Lambda<0}$, and its event horizon is identical to the outermost Killing horizon.
The condition ${f(r_{\rm h})=0}$ gives the relation between the mass parameter ${M}$ and the radius of the Killing horizon ${r_{\rm h}}$ as
\begin{align}
M=-\Lambda r_{\rm h}^2-\kappa Q^2\ln r_{\rm h}=:M_{\rm h}(r_{\rm h}).\label{Mh-def}
\end{align}
In the domain $r\in(0,\infty)$, the function ${M_{\rm h}(r)}$ admits a single local minimum ${M_{\rm ex}:=M_{\rm h}(r_{\rm ex})}$, where 
\begin{align}
&r_{\rm ex}:=\sqrt{\frac{\kappa Q^2}{2(-\Lambda)}},\label{r-ex}\\
&M_{\rm ex}=\frac12\kappa Q^2\biggl\{1-\ln \biggl(\frac{\kappa Q^2}{2(-\Lambda)}\biggl)\biggl\}.\label{extreme-Clement}
\end{align}
As ${r=r_{\rm ex}}$ and ${M=M_{\rm ex}}$ satisfy ${f(r_{\rm ex})=f'(r_{\rm ex})=0}$, ${r=r_{\rm ex}}$ is a degenerate horizon.
For ${M>M_{\rm ex}}$, there are two non-degenerate Killing horizons.
For ${M=M_{\rm ex}}$, there is a single degenerate Killing horizon.
For ${M<M_{\rm ex}}$, there is no horizon.
We note that ${M_{\rm ex}>(<)0}$ holds for ${\kappa Q^2/[2(-\Lambda)]<(>)e}$.

In the uncharged case $Q=0$, in contrast, the spacetime is locally maximally symmetric and the metric (\ref{chargedBTZ}) reduces to
\begin{align}
\label{chargedBTZ-Q=0-F}
\begin{aligned}
&\D s^2=-F\,\D t^2+F^{-1}\D R^2+R^2\biggl(\D\theta-\frac{M\omega}{\sqrt{-\Lambda}\,(1-\omega^2)R^2}\,\D t\biggl)^2,\\
&F(R):=\frac{r^2}{R^2}\,f=-\Lambda R^2-\frac{1+\omega^2}{1-\omega^2}M+\frac{M^2\omega^2}{(-\Lambda)\,(1-\omega^2)^2R^2}
\end{aligned}
\end{align}
in the coordinates ${(t,R,\theta)}$.
The metric (\ref{chargedBTZ-Q=0-F}) is identical to the vacuum rotating BTZ solution (\ref{rotatingBTZ-mj}) with
\begin{align}
&m=\frac{1+\omega^2}{8G(1-\omega^2)}M,\qquad j=\frac{M\omega}{4G\sqrt{-\Lambda}(1-\omega^2)},\label{mj-unchargedClement}
\end{align}
which are solved for ${M}$ and ${\omega}$ to give
\begin{align}
M=&\mp 8G\sqrt{m^2+\Lambda j^2},\qquad \omega=\frac{m\pm\sqrt{m^2+\Lambda j^2}}{j\sqrt{-\Lambda}}. \label{Momega-mj}
\end{align}
The extremality condition (\ref{extreme-mj}) under the parametrization (\ref{rotatingBTZ-mj}) is ${m=\pm \sqrt{-\Lambda}j}$.
Although it gives ${M=0}$ and ${\omega=\pm 1}$ by Eq.~(\ref{Momega-mj}), ${m}$ and ${j}$ are then undetermined in Eq.~(\ref{mj-unchargedClement}).
It shows that the extreme case in vacuum cannot be treated properly under the parametrization (\ref{chargedBTZ-Q=0-F}).
In fact, the only way to give a finite limit is to take ${M\to 0}$ first and next ${\omega^2\to 1}$, but then we have ${g_{t\theta}=0}$ and ${F(R)=-\Lambda R^2}$, which is a particular class of the vacuum BTZ solution (\ref{rotatingBTZ-mj}) with ${m=j=0}$.
For this reason, the uncharged limit ${Q\to 0}$ is not allowed in Eqs.~(\ref{r-ex}) and (\ref{extreme-Clement}).

In the vacuum rotating BTZ solution~(\ref{chargedBTZ-Q=0-F}), ${R=0}$ is not a curvature singularity but just a coordinate singularity, so that the spacetime can be analytically extended beyond there.
This extension is performed by a coordinate transformation ${x:=R^2}$, with which the metric and its inverse are both analytic at ${x=0}$ and the determinant of the metric becomes ${\det g=-1/4}$.
In the extended region ${x<0}$, which corresponds to ${R^2<0}$, there exist closed timelike curves because the squared norm of a Killing vector ${\Theta^\mu=(\partial/\partial\theta)^\mu}$ becomes negative such as ${\Theta_\mu\Theta^\mu=x<0}$.

In the uncharged non-rotating case ${Q=0=\omega}$, in contrast, ${R=0}$ is not always analytic.
For example, the metric (\ref{chargedBTZ-Q=0-F}) with ${M<0}$ and ${\omega=0}$ can be written near ${R=0}$ as
\begin{align}
&\D s^2\simeq -(-M)\,\D t^2+(-M)^{-1}\D R^2+R^2\D\theta^2= -\,\D {\hat t}^2+\D {\hat r}^2+{\hat r}^2\D{\hat \theta}^2,
\end{align}
where ${{\hat t}:=\sqrt{-M}t}$, ${{\hat r}:=R/\sqrt{-M}}$, and ${{\hat\theta}:=\sqrt{-M}\theta}$.
Since the period ${2\pi}$ of ${\theta}$ means the period ${2\pi \sqrt{-M}}$ of ${\hat\theta}$, there is a conical singularity at ${R=0}$ for ${M\ne -1}$.

In the non-rotating limit $\omega\to 0$, the solution (\ref{chargedBTZ})--(\ref{Clement-h}) reduces to the charged non-rotating BTZ solution~\cite{Banados:1992wn,Peldan:1992mp} given by
\begin{align}
\label{chargedBTZ:a=0}
\begin{aligned}
&\D s^2=-f\,\D t^2+f^{-1}\D r^2+r^2\D\theta^2,\\
&A_\mu\D x^\mu=-Q\ln r \, \D t,\\
&f(r)=-\Lambda r^2-M-\kappa Q^2\ln r.
\end{aligned}
\end{align}
It is emphasized that the rotating solution is {\it locally identical} to the non-rotating solution.
In fact, by the following coordinate transformations
\begin{align}
t=\frac{1}{\sqrt{1-\omega^2}}\,{\tilde t}+\frac{\omega}{\sqrt{-\Lambda}\sqrt{1-\omega^2}}\,{\tilde\theta},\qquad \theta=\frac{\omega\sqrt{-\Lambda}}{\sqrt{1-\omega^2}}\,{\tilde t}+\frac{1}{\sqrt{1-\omega^2}}\,{\tilde\theta},\label{local-trans-BTZ}
\end{align}
the rotating solution~(\ref{chargedBTZ})--(\ref{Clement-h}) becomes the non-rotating solution (\ref{chargedBTZ:a=0}) with $t$ and $\theta$ replaced by ${\tilde t}$ and ${\tilde\theta}$.
However, the rotating and non-rotating solutions are globally different if $\theta$ is a periodic coordinate.
In fact, if we identify $(t,r,\theta)$ and $(t,r,\theta+2\pi)$ in the rotating solution (\ref{chargedBTZ}), the transformations (\ref{local-trans-BTZ}) show that $({t},r,{\theta})$ and $({t}+a,r,{\theta}+b)$ are identified in the non-rotating solution (\ref{chargedBTZ:a=0}), where 
\begin{align}
a=-\frac{2\pi \omega}{\sqrt{-\Lambda}\sqrt{1-\omega^2}},\qquad b=\frac{2\pi}{\sqrt{1-\omega^2}}.
\end{align}
In fact, there is another non-rotating limit $\omega\to \pm \infty$.
In this limit, the solution (\ref{chargedBTZ})--(\ref{Clement-h}) reduces to
\begin{align}
\begin{aligned}
&\D s^2=\Lambda r^2\,\D t^2+f^{-1}\D r^2-\frac{f}{\Lambda}\D\theta^2,\\
&A_\mu\D x^\mu=\pm\frac{Q}{\sqrt{\Lambda}}\ln r\,\D\theta,\\
&f(r)=-\Lambda r^2-M-\kappa Q^2\ln r,
\end{aligned}
\end{align}
which is related to the solution (\ref{chargedBTZ:a=0}) by the double Wick rotations $t\to i\theta/\sqrt{-\Lambda}$ and $\theta\to \mp i\sqrt{-\Lambda}t$.

Equations~(\ref{m-rep}) and (\ref{j-rep}) with Eqs.~(\ref{Clement-R})--(\ref{Clement-h}) give the KGBD mass and the KGBD angular momentum of the charged rotating BTZ solution in the Cl\'ement form as
\begin{align}
m=&\frac{1+\omega^2}{8G(1-\omega^2)}(M+8\pi G Q^2\ln r)-\frac{\pi \omega^2Q^2}{1-\omega^2}\biggl(1+\frac{2\pi G Q^2}{\Lambda r^2}\biggl),\label{m-Clement}\\
j=&\frac{\omega}{4G\sqrt{-\Lambda}(1-\omega^2)}\left(M+8\pi G Q^2\ln r-4\pi G Q^2\right).\label{j-Clement}
\end{align}
In the {\it non-rotating} limit $\omega\to 0$, we obtain
\begin{align}
&m=\frac{1}{8G}(M+8\pi GQ^2\ln r),\qquad j=0.
\end{align}
In other {\it non-rotating} limits ${\omega\to \pm\infty}$, we obtain
\begin{align}
m=-\frac{1}{8G}(M+8\pi G Q^2\ln r)+\pi Q^2\biggl(1+\frac{2\pi G Q^2}{\Lambda r^2}\biggl),\qquad j=0.
\end{align}
In the {\it uncharged} limit $Q\to 0$, $m$ and $j$ reduce to the constants given by Eq.~(\ref{mj-unchargedClement}).
In the asymptotically AdS region $r\to \infty$, $m$ and $j$ given by Eqs.~(\ref{m-Clement}) and (\ref{j-Clement}) diverge for $Q\ne 0$ and $\omega Q\ne 0$, respectively.
The divergent terms in $m$ and $j$ are proportional to a scalar product of the gauge field and the generalized Kodama vector (\ref{g-Kodama-general}) defined by 
\begin{align}
\Phi:=-A_\mu K^\mu=\frac{Q}{\sqrt{1-\omega^2}}\ln r.\label{G-potential}
\end{align}

With the metric function ${f(r)}$, Eqs.~(\ref{m-Clement}) and (\ref{j-Clement}) can be written as
\begin{align}
m=&\frac{1+\omega^2}{8G(1-\omega^2)}\biggl(\frac12r f'(r)-f(r)\biggl)+\frac{\pi \omega^2Q^2}{1-\omega^2}\frac{f'(r)}{4\Lambda r}+\frac{\pi Q^2}{2(1-\omega^2)},\label{m-Clement2}\\
j=&\frac{\omega}{4G\sqrt{-\Lambda}(1-\omega^2)}\left(\frac12r f'(r)-f(r)\right).\label{j-Clement2}
\end{align}
While ${m}$ and ${j}$ are constants given by Eq.~(\ref{mj-unchargedClement}) in the uncharged case ${Q=0}$, they depend on ${r}$ for ${Q\ne 0}$.
In particular, on the degenerate horizon ${r=r_{\rm ex}}$ given by Eq.~(\ref{r-ex}) satisfying ${f(r_{\rm ex})=f'(r_{\rm ex})=0}$, the values of ${m}$ and ${j}$ are 
\begin{align}
m(r_{\rm ex})=&\frac{\pi Q^2}{2(1-\omega^2)},\qquad j(r_{\rm ex})=0.\label{mj-Clement-ex}
\end{align}
Unexpectedly, the KGBD quasi-local angular momentum ${j}$ is vanishing on the degenerate horizon in the extremal charged rotating BTZ solution.
Because the present parametrization cannot treat the extreme case properly in the uncharged case, a naive limit ${Q\to 0}$ is not allowed in Eq.~(\ref{mj-Clement-ex}).
In fact, ${j}$ is a non-zero constant in the rotating uncharged case ${Q=0}$ even under the extremality condition ${m=\pm \sqrt{-\Lambda}j}$.

Lastly, using non-zero components of $F^{\mu\nu}$ given by 
\begin{align}
&F^{tr}=-\frac{Q}{\sqrt{1-\omega^2}r},\qquad F^{r\theta}=\frac{\omega Q\sqrt{-\Lambda}}{\sqrt{1-\omega^2}r},
\end{align}
we obtain the dual one-form (\ref{dualF}) for the charged rotating BTZ solution in the Cl\'ement form (\ref{chargedBTZ})--(\ref{Clement-h}) as
\begin{align}
&{}^*F_{t}=\frac{\omega Q\sqrt{-\Lambda}}{\sqrt{1-\omega^2}},\qquad {}^*F_{r}=0,\qquad {}^*F_{\theta}=-\frac{Q}{\sqrt{1-\omega^2}}.
\end{align}
Hence, by $\nabla_\mu\phi\equiv {}^*F_{\mu}$, the expression of the dual massless scalar field $\phi$ is given by 
\begin{align}
\phi=\frac{\omega Q\sqrt{-\Lambda}}{\sqrt{1-\omega^2}}t-\frac{Q}{\sqrt{1-\omega^2}}\theta+\phi_0,
\end{align}
where $\phi_0$ is a constant.
However, different from the Maxwell field, a periodic boundary condition $\phi(t,\theta)=\phi(t,\theta+2\pi)$ cannot be imposed on the dual scalar field.

\subsection{New form for a wider range of $\Lambda$}

In fact, the solution in the Cl\'ement form (\ref{chargedBTZ})--(\ref{Clement-h}) for ${\Lambda<0}$ can be analytically extended for a wider range of $\Lambda$.
By uniform scaling transformations 
\begin{align}
t=\sqrt{1-\omega^2}\,{\bar t},\qquad \theta=\frac{{\bar\theta}}{\sqrt{1-\omega^2}}\label{Clement-New}
\end{align}
with a reparametrization ${a:=\omega/[\sqrt{-\Lambda}\,(1-\omega^2)]}$, the solution is given by the metric (\ref{chargedBTZ}) with
\begin{align}
&A_\mu\D x^\mu=-Q\ln r\,(\D t-a\,\D\theta), \label{new-A}\\
&R(r)=\ \sqrt{\zeta r^2+a^2(M+\kappa Q^2\ln r)},\label{R-def}\\
&f(r)=-\Lambda r^2-M-\kappa Q^2\ln r,\label{f-chargedBTZ-def}\\
&h(r)=-a\,(M+\kappa Q^2\ln r),\label{h-def}
\end{align}
where we have omitted the bars for simplicity, and the constant 
\begin{align}
\zeta=(1-\omega^2)^{-1} \label{zeta-omega}
\end{align}
is determined by
\begin{align}\label{zetaequation}
\zeta^2-\zeta+a^2\Lambda=0,
\end{align}
and hence
\begin{align}
\zeta=\frac12\left(1\pm\sqrt{1-4a^2\Lambda}\right)=:\zeta_\pm.\label{zeta}
\end{align}
If $\theta$ is a periodic coordinate, the rotating solution after the transformations (\ref{Clement-New}) is globally different from the original rotating solution in the Cl\'ement form since an identification of $\theta$ and $\theta+2\pi$ in the latter implies an identification of ${\bar\theta}$ and ${\bar\theta}+2\pi\sqrt{1-\omega^2}$ in the former.

The solution in this new form is parametrized by $M$, $Q$, and $a$, and it is valid for $\Lambda\in(-\infty,\Lambda_{\rm c}]$, where ${\Lambda_{\rm c}:=1/(4a^2)}$ \,(${\,>0}$). 
Since the form of the function $f(r)$ (\ref{Clement-f}) remains the same, the location of the Killing horizon $r=r_{\rm h}$ is unchanged, and ${R^2(r_{\rm h})=\zeta^2r^2_{\rm h}}$ is satisfied.
Hence, the reality condition $R^2>0$ is always satisfied at $r=r_{\rm h}$.
In fact, the solution in the new form with $Q\ne 0$ and $\Lambda\in [0,\Lambda_{\rm c}]$ admits a single Killing horizon for any value of $M$.
However, consistent with Ida's no-go theorem~\cite{Ida:2000jh}, the solution does {\it not} describe a black hole in that case because the trapped region is then given by $r>r_{\rm h}$, so that the Killing horizon is not the outer boundary of the trapped region.

For ${\Lambda< \Lambda_{\rm c}}$, there are two real branches of solutions. They coincide for ${\Lambda=\Lambda_{\rm c}}$, while the metric becomes complex and unphysical for ${\Lambda>\Lambda_{\rm c}}$. We refer to the solution with $\zeta_+$ ($\zeta_-$) as the plus-branch (minus-branch) solution, and both branches satisfy $\lim_{r\to \infty}R^{\mu\nu}_{~~\rho\sigma}=\Lambda(\delta^\mu_{~\rho}\delta^\nu_{~\sigma}-\delta^\mu_{~\sigma}\delta^\nu_{~\rho})$.
\begin{table}[htb]
\begin{center}
\vskip 3mm
\caption{\label{table:Solutions} Asymptotic behavior as ${r\to \infty}$ of the charged rotating BTZ solution (\ref{chargedBTZ}) in different forms depending on the value of $\Lambda$. Here ``dS'', ``flat'', and ``AdS'' stand for asymptotically locally de~Sitter, flat, and anti-de~Sitter spacetime, respectively.}
\vskip 3mm
\scalebox{1.0}{
\begin{tabular}{|c|c|c|c|c|c|c|c|}
\hline
Forms & $\Lambda<0$ & $\Lambda=0$ & $\Lambda\in (0,\Lambda_{\rm c}]$ & $\Lambda>\Lambda_{\rm c}$ \\ \hline\hline
Plus-branch & AdS & flat & dS & n.a. \\ \hline
Minus-branch & unphysical & flat & dS & n.a. \\ \hline
Cl\'ement & AdS & n.a. & n.a. & n.a. \\ \hline
\hline
\end{tabular}
}\end{center}
\end{table}

For ${\Lambda=0}$, we obtain ${\zeta_+=1}$ and ${\zeta_-=0}$, and the two branches are both asymptotically locally flat as ${r\to \infty}$. For ${0<\Lambda\le \Lambda_{\rm c}}$, the two branches are both asymptotically locally dS as $r\to \infty$.
For $\Lambda<0$, we have $\zeta_+>0$ and $\zeta_-<0$, so that the metric in the minus-branch becomes complex in the asymptotically locally AdS region $r\to \infty$. Those properties are summarized in Table~\ref{table:Solutions}.
If the parameters admit real solutions to ${f(r)=0}$ for ${\Lambda<0}$ with ${\zeta=\zeta_+}$, the metric describes an asymptotically AdS charged rotating black hole with an event horizon at ${r=r_{\rm EH}}$, where $r_{\rm EH}$ is the largest root of ${f(r)=0}$.

Now we consider {\it three special cases} of the solution in the new form given by the metric (\ref{chargedBTZ}) with Eqs.~(\ref{new-A})--(\ref{h-def}) for ${M\ne0}$. In the {\it uncharged} case (${Q=0}$) for ${\zeta\ne 0}$, using $R$ as a radial coordinate, one can write both branches of solutions in the same form as
\begin{align}
\label{chargedBTZ:Q=0}
\begin{aligned}
&\D s^2=-F\,\D {\bar t}\,^2+F^{-1}\D R^2+R^2\biggl(\D\theta-\frac{{\bar a}}{R^2}\D {\bar t}\biggl)^2,\\
&A_\mu\D x^\mu=0,\\
&F(R)=-\Lambda R^2-{\bar M}+\frac{{\bar a}^2}{R^2},
\end{aligned}
\end{align}
where ${{\bar t}:=t/\zeta}$ and
\begin{align}
{\bar M}:=\zeta(2\zeta-1)\,M,\qquad {\bar a}:=\zeta\,aM.
\end{align}
This is the uncharged rotating BTZ solution parametrized by ${\bar M}$ and ${\bar a}$~\cite{Banados:1992wn} which is locally maximally symmetric. For ${\zeta=0}$, the limit ${Q\to 0}$ is possible only for ${\Lambda=0}$ with ${a\ne 0}$ in the minus-branch by Eq.~(\ref{zetaequation}), and then the solution reduces to the following locally flat spacetime
\begin{align}
\D s^2=-\D {\bar t}^2+{\bar t}^2\,\D {\bar r}^2+\D{\bar\theta}^2,
\end{align}
where ${{\bar r}:=(\sqrt{M}/a)\,t}$, ${{\bar t}:=r/\sqrt{M}}$, and ${{\bar \theta}:=\sqrt{M}\,(a\,\theta-t)}$.

In the case without a cosmological constant (${\Lambda=0}$), the solution reduces to
\begin{align}
\label{chargedBTZ:Lambda=0}
\begin{aligned}
&\D s^2=-\frac{r^2}{R^2}\,f\,\D t^2+f^{-1}\D r^2+R^2\biggl(\D\theta+\frac{h}{R^2}\D t\biggl)^2,\\
&A_\mu\D x^\mu=-Q\ln r\,(\D t-a\,\D\theta),\\
&f(r)=-M-\kappa Q^2\ln r,
\end{aligned}
\end{align}
with $h(r)$ and $R(r)$ given by Eqs.~(\ref{h-def}) and (\ref{R-def}), where ${\zeta=1\,(=\zeta_+)}$ or ${\zeta=0\,(=\zeta_-)}$.
In the minus-branch (${\zeta=0}$), by coordinate transformations ${\hat t}:=t-a\,\theta$ and ${\hat \theta}:=t/a$, the solution is written as
\begin{align}
\begin{aligned}
&\D s^2=-(-M-\kappa Q^2\ln r)\,\D {\hat t}^2+(-M-\kappa Q^2\ln r)^{-1}\,\D r^2+r^2\,\D {\hat \theta}^2,\\
&A_\mu\D x^\mu=-Q\ln r \,\D {\hat t},
\end{aligned}
\end{align}
which is the non-rotating charged BTZ solution with ${\Lambda=0}$.
(See also the electrostatic solution given by Eq. (11.54) in Sec.~11.2.2 of~\cite{Garcia-Diaz:2017cpv}.)

Lastly, only the positive-branch solution admits the {\it non-rotating limit} ${a\to 0}$ (implying ${\zeta(\zeta-1)=0}$ so that only the case ${\zeta=1}$ is possible) given by Eq.~(\ref{chargedBTZ:a=0}).
In contrast, the negative-branch solution does not admit a non-rotating limit ${a\to 0}$ due to ${\lim_{a \to 0}R=0}$.

Equations~(\ref{m-rep}) and (\ref{j-rep}) with Eqs.~(\ref{R-def})--(\ref{h-def}) give the KGBD mass and the KGBD angular momentum of the charged rotating BTZ solution in the new form as
\begin{align}
m=&\frac{\zeta(2\zeta-1)}{8G}(M+8\pi G Q^2\ln r)+\pi a^2 Q^2\biggl(\Lambda+ \frac{2\pi G Q^2}{r^2}\biggl),\label{m-newform} \\
j=&\frac{\zeta a}{4G}\left(M+8\pi G Q^2\ln r-4\pi G Q^2\right).\label{j-newform}
\end{align}
The expressions (\ref{m-newform}) and (\ref{j-newform}) are consistent with Eqs.~(\ref{m-Clement}) and (\ref{j-Clement}) for the Cl\'ement form because the transformations (\ref{Clement-New}) with Eq.~(\ref{zeta-omega}) put a constant factor $\sqrt{1-\omega^2}$ on the axial Killing vector $\psi^\mu$.
The quantities $m$ and $j$ defined by Eqs.~(\ref{def-QLM2}) and (\ref{def-QLJ2}), respectively, are quadratic of $\psi^\mu$.

In the {\it non-rotating} limit $a\to 0$ (and then only $\zeta=\zeta_+=1$ is possible), we obtain
\begin{align}
&m=\frac{1}{8G}(M+8\pi G Q^2\ln r),\qquad j=0.
\end{align}
In the {\it uncharged} limit $Q\to 0$, $m$ and $j$ reduce to the constants
\begin{align}
&m=\frac{\zeta(2\zeta-1)}{8G}M,\qquad j=\frac{\zeta }{4G}aM,
\end{align}
which give $m=j=0$ for $\zeta=0$ (realized only for $\Lambda=0$) and $m=0$ in the degenerate case $\zeta=1/2$ for $a^2\Lambda=1/4$. 
In the asymptotic region $r\to \infty$, $m$ given by Eq.~(\ref{m-newform}) diverges for $\zeta(2\zeta-1)Q\ne 0$, while $j$ given by Eq.~(\ref{j-newform}) diverges for $\zeta aQ\ne 0$.
Therefore in the charged case $Q\ne 0$, both $m$ and $j$ are finite as $r\to \infty$ only for $\zeta=0$ with $\Lambda=0$.

\subsection{New Doran-type horizon-penetrating coordinates}

In this subsection, we present a new coordinate system of the charged rotating BTZ solution which covers the Killing horizon.
By the coordinate transformations
\begin{align}
&\D t=\D T+\epsilon_r \frac{\sqrt{R^2-r^2f}}{rf}\,\D r,\label{T-def2}\\
&\D\theta=\D\varphi-\epsilon_r\frac{h}{rf\sqrt{R^2-r^2f}}\,\D r, \label{varphi-def2}
\end{align}
with ${\epsilon_r=\pm 1}$, the metric ~(\ref{chargedBTZ}) with {\it arbitrary} $R(r)$, $f(r)$, and $h(r)$ is written in the {\it new coordinate system} $(T,r,\varphi)$ as
\begin{align}
&\D s^2=-\D T^2+\frac{R^2(R^2-r^2f)}{R^2 - r^2f + h^2}\,\D\varphi^2 \nonumber\\
&\hspace{12mm}
+\frac{r^2(R^2 - r^2f + h^2)}{R^2(R^2-r^2f)}\bigg[\D r-\epsilon_r\frac{\sqrt{R^2- r^2f}}{r}\biggl(\D T+\frac{hR^2}{R^2 - r^2f + h^2}\,\D\varphi\biggl)\bigg]^2. \label{BTZ-Doran}
\end{align}
The derivation of the new coordinate system is presented in Appendix~\ref{app:derivation}. 
The determinant of the metric (\ref{BTZ-Doran}) in the new coordinates is ${\det g=-r^2}$, while the metric and its inverse are given by
\begin{align}
\begin{aligned}
&g_{TT}=-\frac{r^2f - h^2}{R^2},\qquad g_{Tr}=-\frac{\epsilon_r r(R^2 - r^2f + h^2)}{R^2\sqrt{R^2 - r^2f}},\qquad g_{T\varphi}=h,\\
&g_{rr}=\frac{r^2(R^2 - r^2f + h^2)}{R^2(R^2-r^2f)},\qquad g_{r\varphi}=-\frac{\epsilon_r rh}{\sqrt{R^2- r^2f}},\qquad g_{\varphi\varphi}=R^2,
\end{aligned}
\end{align}
and
\begin{align}
\begin{aligned}
&g^{TT}=-1,\qquad g^{Tr}=-\frac{\epsilon_r\sqrt{R^2-r^2f}}{r},\qquad g^{T\varphi}=0,\\
&g^{rr}=f,\qquad g^{r\varphi}=\frac{\epsilon_rh}{r\sqrt{R^2-r^2f}},\qquad g^{\varphi\varphi}=\frac{R^2 - r^2f + h^2}{R^2(R^2-r^2f)},
\end{aligned}
\end{align}
respectively, which are regular on the horizon ${f(r_{\rm h})=0}$.
Two remarkable properties of the new coordinate system (\ref{BTZ-Doran}) are as follows:
\begin{itemize}
\item \ ${g^{TT}=-1}$.

\item \ The time coordinate ${T}$ coincides with the proper time along timelike geodesics with ${E=1}$ and ${L=0}$. (See Appendix~\ref{app:derivation}.)
\end{itemize}
\noindent
Therefore, our new coordinate system is the counterpart of the Doran coordinates in the Kerr spacetime~\cite{Doran:1999gb} which share the same properties.
The coordinate system (\ref{BTZ-Doran}) was obtained in~\cite{Zaslavskii:2018lbb} with particular forms of $R(r)$, $f(r)$, and $h(r)$ for the uncharged rotating BTZ solution.
As the new coordinate system is regular on the Killing horizon ${r=r_{\rm h}}$, the surface gravity ${\kappa_{\rm h}}$ on the horizon can be computed from the definition ${\xi^\nu\nabla_\nu \xi^\mu=\kappa_{\rm h} \xi^\mu}$ evaluated at ${r=r_{\rm h}}$. 
With the associated Killing vector ${\xi^\mu=(1,0,-h(r_{\rm h})/R^2(r_{\rm h}))}$ in the coordinate system (\ref{BTZ-Doran}), we obtain
\begin{align}
\kappa_{\rm h}=&-\epsilon_r\frac{rf'}{2R}\biggl|_{r=r_{\rm h}}.
\end{align}

In order for the metric to be real, the new coordinates (\ref{BTZ-Doran}) cover the original spacetime (\ref{chargedBTZ}) only in the domain where $R^2> r^2f$ holds.
Consequently, we can identify a {stably causal} region in the spacetime described by the metric (\ref{chargedBTZ}).
A time-orientable general spacetime is said to be {\it causal} if there is no closed causal curve~\cite{Hawking:1973uf} and {\it stably causal} if it is causal and no closed causal curve appears even under any small perturbation against the metric.
\begin{Prop}
\label{Prop:Causal}
A spacetime described by the metric (\ref{chargedBTZ}) is stably causal in the region where $R^2> r^2f$ holds.
\end{Prop}
{\it Proof}.
By Proposition 6.4.9 in~\cite{Hawking:1973uf}, a time-orientable spacetime is everywhere stably causal if and only if there is a {\it time function} ${\cal T}$, which is a differentiable function giving timelike $\nabla_\mu {\cal T}$.
Since $R^2> r^2f$ holds by assumption, we can use the coordinates (\ref{BTZ-Doran}).
Then, since a vector $U^\mu:=\nabla^\mu T$ is everywhere timelike satisfying $U_\mu U^\mu=-1$, the spacetime (\ref{BTZ-Doran}) is time-orientable by $U^\mu$.
Furthermore, since $T$ is a time function, the spacetime is stably causal.
\qed

Different from the single-null coordinates (\ref{null-coord}), the charged rotating BTZ solution is described by {\it elementary functions} in the new coordinates (\ref{BTZ-Doran}) with Eqs.~(\ref{new-A})--(\ref{h-def}). As we have ${R^2- r^2f=\zeta r^2+(r^2+a^2)(M+\kappa Q^2\ln r)+\Lambda r^4}$, the new coordinates (\ref{BTZ-Doran}) do {\it not} cover the asymptotically AdS region ${r\to \infty}$ for ${\Lambda<0}$ and near the singularity ${r=0}$ for any ${\Lambda\in(-\infty,\Lambda_{\rm c}]}$.
Nevertheless, they cover the Killing horizon ${r=r_{\rm h}}$, as the metric and its inverse are both regular. The gauge field is written in the new coordinates as
\begin{align}
&A_\mu\D x^\mu=-Q\ln r\left(\D T+\epsilon_r\frac{R^2-r^2f+ah}{rf\sqrt{R^2-r^2f}}\,\D r-a\,\D\varphi\right).\label{A-Doran}
\end{align}
It is seen that $A_r$ diverges on the horizon, but we can always set $A_r=0$ by a gauge transformation $A_\mu\to A_\mu+\nabla_\mu B$ with an appropriate $B=B(r)$. 
In fact, non-zero components of the Faraday tensor are given by
\begin{align}
F_{Tr}=-F_{rT}=\frac{Q}{r},\qquad F_{r\varphi}=-F_{\varphi r}=a\,\frac{Q}{r},
\end{align}
which are finite at the Killing horizon, and give ${F_{\mu\nu}F^{\mu\nu}=-2Q^2/r^2}$ in agreement with Eq.~(\ref{Clement-F2}). 
With ${a=0}$, which is allowed only in the plus branch (and then ${\zeta=1}$ so that $R=r$), the solution reduces to the charged non-rotating BTZ solution in the Painlev\'{e}-Gullstrand coordinates,
\begin{align}
\label{BTZ-PG}
\begin{aligned}
&\D s^2=-\D T^2+ r^2\,\D\varphi^2+\left(\D r-\epsilon_r\sqrt{1-f}\,\D T\right)^2, \\
&A_\mu\D x^\mu=-Q\ln r\left(\D T+\epsilon_r\frac{\sqrt{1-f}}{f}\,\D r\right),
\end{aligned}
\end{align}
where $f(r)$ is given by Eq.~(\ref{f-chargedBTZ-def}).

\subsection{Cotton and Hawking-Ellis types}

Here we clarify the algebraic structures of the charged rotating BTZ solution. Since the solution is locally identical to the non-rotating solution and the algebraic structures are invariant under coordinate transformations, it is sufficient to study the simpler charged non-rotating BTZ solution (\ref{chargedBTZ:a=0}).

First, we study the Cotton type of the charged non-rotating BTZ spacetime.
Recently, a novel and more practical method of classification of spacetimes in three-dimensional gravity, consistent with \cite{Garcia:2003bw}, was suggested \cite{Podolsky:2023qiu,Papajcik:2023zen}. It uses five real Cotton scalars~$\Psi_{\rm A}$ defined by the following direct projections of the {\it Cotton tensor} $C_{\mu\nu\rho}$~\cite{Cotton:1899} onto a properly normalized null triad ${\{k^\mu, l^\mu, m^\mu\}}$,
\begin{align}\label{Cottonscalars}
\begin{aligned}
&\Psi_0:=C_{\mu\nu\rho}\,k^\mu m^\nu k^\rho, \\
&\Psi_1:=C_{\mu\nu\rho}\,k^\mu l^\nu k^\rho,\\
&\Psi_2:=C_{\mu\nu\rho}\,k^\mu m^\nu l^\rho,\\
&\Psi_3:=C_{\mu\nu\rho}\,l^\mu k^\nu l^\rho, \\
&\Psi_4:=C_{\mu\nu\rho}\,l^\mu m^\nu l^\rho,
\end{aligned}
\end{align}
which are three-dimensional counterparts of the famous Newman-Penrose complex Weyl scalars of four-dimensional gravity~\cite{Newman:1961qr}.
Here the Cotton tensor is defined as (following the convention of~\cite{Garcia:2003bw})
\begin{align}
C_{\mu\nu\rho}:=2\,\big(\nabla_{[\mu}R_{\nu]\rho}-\tfrac14\nabla_{[\mu}R\,g_{\nu]\rho}\big),
\end{align}
which automatically satisfies the constraints ${C_{(\mu\nu)\rho}\equiv 0}$, ${C_{[\mu\nu\rho]}\equiv 0}$, and ${C_{\mu\nu}^{~~\mu}\equiv 0}$, and $k^\mu$, $l^\mu$, and $m^\mu$ are the basis vectors of the spacetime satisfying
\begin{align}\label{triad}
\begin{aligned}
&k_\mu k^\mu=0=l_\mu l^\mu, \qquad k_\mu l^\mu=-1, \\
&m_\mu m^\mu=1, \qquad k_\mu m^\mu=0=l_\mu m^\mu.
\end{aligned}
\end{align}

The locally trivial type~O geometry is a conformally flat spacetime with vanishing Cotton tensor (if and only if $C_{\mu\nu\rho}\equiv 0$). It occurs in the case when ${\Psi_A=0}$ holds for all $A$.
The uncharged (${Q=0}$) rotating BTZ spacetime is locally AdS and therefore conformally flat, so that it is of type O everywhere. In contrast, the charged (${Q\ne0}$) rotating BTZ spacetime is not conformally flat, so that its Cotton type is non-trivial. 
To determine the Cotton type of the charged rotating BTZ spacetime, we will use the following proposition.
\begin{Prop}
\label{Prop:Cotton-general}
Consider the most general static and circularly symmetric spacetime described by the metric in the single-null coordinates $(v,x,\theta)$ given by 
\begin{align}
\label{single-null}
&\D s^2=-f\,\D v^2+2\,\D v\D x+r^2\D\theta^2
\end{align}
with $f=f(x)$ and $r=r(x)$, and assume that the metric functions $f$ and $r$ are at least ${C^{2,1}}$ (often denoted by ${C^{3-}}$ in physics)\footnote{The metric functions are at least $C^2$ and have the second derivatives that are locally Lipschitz continuous, which restricts the third derivatives to be finite but allows their finite jumps.}.
Then, the Cotton type of the spacetime in a region with $r\ne 0$ is determined as follows:
\begin{align}
\begin{aligned}
&2f(rr_{,xxx}-r_{,x}r_{,xx}) - r(rf_{,xxx}-4f_{,x}r_{,xx})=0~~\to~~\mbox{\rm type~O}, \\
&2f(rr_{,xxx}-r_{,x}r_{,xx}) - r(rf_{,xxx}-4f_{,x}r_{,xx})\ne 0~~\mbox{\rm with}~~f\ne 0~~\to~~\mbox{\rm type~I}, \\
&2f(rr_{,xxx}-r_{,x}r_{,xx}) - r(rf_{,xxx}-4f_{,x}r_{,xx})\ne 0~~\mbox{\rm with}~~f= 0~~\to~~\mbox{\rm type~III}.
\end{aligned}
\end{align}
\end{Prop}
{\it Proof}.
For the basis one-forms given by
\begin{align}
k_\mu\D x^\mu&=-\D v, \nonumber\\ 
l_\mu\D x^\mu&=-\frac12f\D v+\D x,\label{l-1-general}\\
m_\mu\D x^\mu&=r\,\D{\theta}, \nonumber 
\end{align}
we obtain $k^\nu \nabla_\nu k^\mu=0$, and therefore $k^\mu$ is tangent to a null geodesic.
The optical scalars~\cite{Podolsky:2018zha} in three dimensions for $k^\mu$ and $l^\mu$ are computed to give
\begin{align}
\rho_k&:=(\nabla_\mu k_{\nu})\,m^\mu m^\nu=-\frac{r_{,x}}{r},\\
\rho_l&:=(\nabla_\mu l_{\nu})\,m^\mu m^\nu=\frac{fr_{,x}}{2r}.
\end{align}
Note that the basis one-forms (\ref{l-1-general}) and their inverses are finite with $f=0$.
Then, the Cotton scalars (\ref{Cottonscalars}) are given by 
\begin{align}
\label{C-scalar-g}
\begin{aligned}
&\Psi_0=\Psi_2=\Psi_4=0, \\
&\Psi_1=\frac{1}{4r^2}\left[2f(rr_{,xxx}-r_{,x}r_{,xx}) - r(rf_{,xxx}-4f_{,x}r_{,xx})\right], \\
&\Psi_3=-\frac12f\Psi_1.
\end{aligned}
\end{align}
If $f$ and $r$ are at least $C^{2,1}$, the Cotton scalars are finite with $r\ne 0$.
Then, as ${\Psi_0=0=\Psi_4}$ holds, we can use Table~II in~\cite{Podolsky:2023qiu} (which is Table~IV in~\cite{Papajcik:2023zen}) to determine the algebraic type of the spacetime, according to the structure of its Cotton tensor.
The table shows that the spacetime is of the Cotton type O if ${\Psi_1=0=\Psi_3}$, type I if ${\Psi_1\ne 0}$ and ${\Psi_3\ne 0}$, and type III if ${\Psi_1\ne 0}$ and ${\Psi_3= 0}$, so that the proposition follows from Eq.~(\ref{C-scalar-g}).
\qed

Proposition~\ref{Prop:Cotton-general} shows the Cotton type of the charged non-rotating BTZ solution as follows.
\begin{Coro}
\label{Coro:Cotton}
The uncharged rotating BTZ spacetime (with ${Q=0}$) is of the Cotton type~O everywhere.
The charged rotating BTZ spacetime (with ${Q\ne0}$) is of the Cotton type I at $r\ne r_{\rm h}$, and of the Cotton type III on the Killing horizon $r=r_{\rm h}$.
\end{Coro}
{\it Proof}.
The charged non-rotating BTZ solution (\ref{chargedBTZ:a=0}) is described by the metric (\ref{single-null}) with ${f(x)=-\Lambda x^2-M-\kappa Q^2\ln x}$ and ${r(x)=x}$, which gives
\begin{align}
2f(rr_{,xxx}-r_{,x}r_{,xx}) - r(rf_{,xxx}-4f_{,x}r_{,xx})=\frac{2\kappa Q^2}{x}.
\end{align}
Then, the corollary follows from Proposition~\ref{Prop:Cotton-general}.
\qed

\noindent
Corollary~\ref{Coro:Cotton} is consistent with the book~\cite{Garcia-Diaz:2017cpv}, in which the charged rotating BTZ solution is shown to be of the Cotton type~I {\it away from the horizon} in Sec.~11.11.2 with the Cl\'ement form~\cite{Clement:1995zt}, and in Sec.~11.11.3 with the Mart\'{\i}nez-Teitelboim-Zanelli form~\cite{Martinez:1999qi}.

Finally, we study the Hawking-Ellis type of the energy-momentum tensor for the Maxwell field.
(See Sec.~3 in~\cite{Maeda:2018hqu} for the Hawking-Ellis classification in arbitrary dimensions.)
{It is almost a trivial task to confirm that Propositions 1 and 2 in~\cite{Maeda:2021ukk} remain valid in the presence of $\Lambda$.
Therefore, the energy-momentum tensor of the Maxwell field in the solution (\ref{chargedBTZ:a=0}) is of the Hawking-Ellis type I everywhere, including the Killing horizon $r=r_{\rm h}$.
\begin{Prop}
\label{Prop:HE}
The energy-momentum tensor of the Maxwell field in the charged rotating BTZ solution is of the Hawking-Ellis type I everywhere, including the Killing horizon ${r=r_{\rm h}}$.
\end{Prop}

\section{Summary}
\label{sec:summary}

In this paper, we have revisited the charged rotating BTZ solution in the three-dimensional Einstein-Maxwell-$\Lambda$ system.
Our main results can be summarized as follows.

\begin{enumerate}
\item We have extended the charged rotating BTZ solution in the Cl\'ement form (\ref{chargedBTZ})--(\ref{Clement-h}) for ${\Lambda<0}$ to a wider range of $\Lambda$. Our new form is given by the metric~(\ref{chargedBTZ}) with Eqs.~(\ref{new-A})--(\ref{h-def}), and valid for any ${\Lambda\le 1/(4a^2)}$. The new form consists of two branches, and the metric in the plus-branch is real and physical in the asymptotically AdS region ${r\to \infty}$ for ${\Lambda<0}$.

\item We have presented the new coordinates (\ref{BTZ-Doran}) nicely covering the Killing horizon ${r=r_{\rm h}}$ defined by ${f(r_{\rm h})=0}$. The new coordinates cover the original spacetime (\ref{chargedBTZ}) only in the region where ${R^2> r^2f}$ holds. As a consequence, by Proposition~\ref{Prop:Causal}, the region given by $R^2> r^2f$ in a spacetime described by the metric (\ref{chargedBTZ}) --- not necessarily to be the charged rotating BTZ spacetime --- is stably causal.

\item We have computed the KGBD mass $m$ and the KGBD angular momentum $j$ for the charged rotating BTZ solution in the Cl\'ement form, and also in the new form as Eqs.~(\ref{m-Clement}) and (\ref{j-Clement}) and Eqs.~(\ref{m-newform}) and (\ref{j-newform}), respectively.
They are finite in the uncharged case $Q=0$, but for $Q\ne 0$ at least either of them diverges in the asymptotic region $r\to\infty$, with the exception of $\zeta=0$ (for $\Lambda=0$) in the new form. 
In particular, ${j=0}$ is satisfied on the degenerate horizon ${r=r_{\rm ex}}$ in the charged rotating case ${Q\ne 0}$ in spite that ${j}$ is non-zero in the uncharged rotating case ${Q=0}$ even under the extremality condition ${m=\pm \sqrt{-\Lambda}j}$.

\item By Corollary~\ref{Coro:Cotton}, the uncharged rotating BTZ spacetime (${Q=0}$) is of the Cotton type~O (that is conformally flat) everywhere. The charged rotating BTZ spacetime (${Q\ne0}$) is of the Cotton type I at ${r\ne r_{\rm h}}$, and of the Cotton type III on the Killing horizon ${r=r_{\rm h}}$. Interestingly, this is analogous to the charged static black holes in the Robinson-Trautman form presented in Sec.~VIII of \cite{Podolsky:2023qiu}.

\item By Proposition~\ref{Prop:HE}, the energy-momentum tensor of the Maxwell field in the charged rotating BTZ solution is of the Hawking-Ellis type I everywhere, including the Killing horizon.

\end{enumerate}

Related to the result 3, as seen in Eqs.~(\ref{m-Clement}), (\ref{j-Clement}), and (\ref{G-potential}), the KGBD quasi-local mass and angular momentum as well as the electric potential measured by the Kodama observer diverge as $r\to \infty$ for the charged rotating BTZ solution.
Nevertheless, finite values of the global mass, angular momentum, and electric charge have been obtained as the Regge-Teitelboim charges based on the Hamiltonian formalism of the Einstein-Maxwell-$\Lambda$ system~\cite{Martinez:1999qi}.
The relation between those quasi-local quantities and the global quantities for the charged rotating BTZ black hole is an important open question.

The results 4 and 5 show that algebraic properties on the Killing horizon are quite different from other regions of the spacetime.
In comparison, it was shown in four dimensions~\cite{Maeda:2021jdc} that a spacetime described by the G\"urses-G\"ursey metric~\cite{Gurses:1975vu} in the Doran coordinates ${(T,r,\theta,\varphi)}$,
\begin{align}
\D s^2=&-\D T^2+(r^2+a^2\cos^2\theta)\,\D \theta^2+(r^2+a^2)\sin^2\theta\,\D\varphi^2\nonumber \\
&+\frac{r^2+a^2\cos^2\theta}{r^2+a^2}\bigg[\D r+\frac{\sqrt{2M(r)\,r\,(r^2+a^2)}}{r^2+a^2\cos^2\theta}(\D T-a\sin^2\theta\,\D\varphi)\bigg]^2,\label{Doran-GG}
\end{align}
with an arbitrary mass function $M(r)$ satisfying ${rM(r)\ge 0}$ is of the Petrov type D {\it everywhere} and the corresponding energy-momentum tensor in general relativity is of the Hawking-Ellis type I {\it everywhere} including the Killing horizon ${r=r_{\rm h}}$ defined by ${\Delta(r_{\rm h})=0}$ with ${\Delta(r)=r^2-2rM(r)+a^2}$.
The metric (\ref{Doran-GG}) satisfies ${g^{TT}=-1}$, and includes the Kerr spacetime for ${M(r)=M_0}$ and the Kerr-Newman spacetime for ${M(r)=M_0-Q^2/r}$ as special cases, where $M_0$ and $Q$ are constants.
Such rigorous algebraic classifications of spacetime geometry and of the energy-momentum tensor on the horizon in a more general spacetime, and in higher dimensions, are still open. 
We leave these problems for future investigation.

\section*{Acknowledgments}
The authors thank Cristi{\'a}n Mart\'{\i}nez for valuable communications on the charged rotating BTZ solution. They also thank the anonymous referees for their helpful comments that improved the quality of the manuscript considerably. The authors are very grateful to the Max-Planck-Institut f\"ur Gravitationsphysik (Albert-Einstein-Institut) for its hospitality. H.M. is also grateful to the Institute of Theoretical Physics of the Charles University, Prague, where this work started, for its hospitality. This work has been supported by the Czech Science Foundation Grant No.~GA\v{C}R 23-05914S.

\appendix

\section{Cl\'ement's solution in the 1993 paper}
\label{app:Clement}

In this appendix, we show that Cl\'ement's solution given by Eq.~(23) in~\cite{Clement:1993kc} is locally identical to the charged non-rotating BTZ solution.
The original solution is written in the coordinates $({t},\rho,{\theta})$ as
\begin{align}
\begin{aligned}
&\D s^2=U(\D {t}-\omega\,\D\theta)^2+W\D{\theta}^2+\frac{\D\rho^2}{2\rho U},\\
&A_\mu\D x^\mu=-q\ln\biggl(\frac{\rho}{\rho_0}\biggl)(\D {t}-\omega\,\D{\theta}),\\
&U(\rho)=2\biggl(-\Lambda\rho+\kappa q^2\ln\biggl(\frac{\rho}{\rho_0}\biggl)\biggl),\qquad W(\rho)=-2\rho,
\end{aligned}
\end{align}
where $\rho_0$ and $q$ are parameters.
By coordinate transformations ${{t}-\omega\,{\theta}=i\,{\bar t}}$, ${\rho=r^2/2}$, and ${{\theta}=i\,{\bar\theta}}$ together with reparametrizations ${Q:=2iq}$ and ${M:=2\kappa q^2\ln(2\rho_0)}$, the solution becomes
\begin{align}
\begin{aligned}
&\D s^2=-f\,\D {\bar t}^2+f^{-1}\D r^2+r^2\D{\bar\theta}^2,\\
&A_\mu\D x^\mu=-Q\ln r\,\D {\bar t}+\mbox{constant},\\
&f(r)=-\Lambda r^2-M-\kappa Q^2\ln r,
\end{aligned}
\end{align}
which is Eq.~(\ref{chargedBTZ:a=0}).

\section{Derivation of the Doran-type coordinates}
\label{app:derivation}

In this appendix, we present how to systematically find the transformations (\ref{T-def2}) and (\ref{varphi-def2}) from the original coordinates (\ref{chargedBTZ}) to the Doran-type coordinates (\ref{BTZ-Doran}). For derivation, we consider an affinely parametrized geodesic $\gamma$ with its tangent vector ${v^\mu\,(=\D x^\mu/\D\lambda)}$, where $\lambda$ is an affine parameter along $\gamma$ in the original coordinate system (\ref{chargedBTZ}). As the spacetime (\ref{chargedBTZ}) in the coordinates $(t,r,\theta)$ admits two Killing vectors ${\xi^\mu=(1,0,0)}$ and ${\Theta^\mu=(0,0,1)}$, there are conserved quantities ${E:=-\xi_\mu v^\mu}$ and ${L:=\Theta_\mu v^\mu}$ along $\gamma$. Using them and the normalization ${\varepsilon=v_\mu v^\mu}$, where ${\varepsilon=-1}$, $0$, and $1$, corresponds to timelike, null, and spacelike $\gamma$, respectively, we write down the geodesic equations as
\begin{align}
{\dot t}=&\ \frac{ER^2+Lh}{r^2f},\label{geo-BTZ1}\\
{\dot r}=&\ \epsilon_r\frac{\sqrt{(ER^2+Lh)^2 + r^2f(\varepsilon R^2 - L^2)}}{rR},\label{geo-BTZ2}\\
{\dot \theta}=&\ \frac{Lr^2f-h(ER^2+Lh)}{r^2R^2f}.\label{geo-BTZ3}
\end{align}
where ${\epsilon_r=1\,(-1)}$ corresponds to an outgoing (ingoing) $\gamma$.

Equations~(\ref{geo-BTZ2}) and (\ref{geo-BTZ3}) give
\begin{align}
&\frac{\D\theta}{\D r}=\frac{{\dot \theta}}{\dot r}=\gamma(r),\label{dtheta}
\end{align}
where
\begin{align}
&\gamma(r):=\epsilon_r\frac{Lr^2f-h(ER^2+Lh)}{rRf\sqrt{(ER^2+Lh)^2 + r^2f(\varepsilon R^2 - L^2)}}.\label{gamma}
\end{align}
Equation~(\ref{dtheta}) shows that a new coordinate $\varphi$ defined by
\begin{align}
&\D\varphi=\D\theta- \gamma(r)\,\D r\label{varphi-def}
\end{align}
is constant along $\gamma$.

Now consider a new time coordinate $T$ given by
\begin{align}
\D T=\D t+\beta(r)\,\D r, \label{T-def}
\end{align}
and impose ${{\dot T}=1}$ along $\gamma$.
Then, using Eqs.~(\ref{geo-BTZ1}) and (\ref{geo-BTZ2}), we obtain $\beta(r)$ as
\begin{align}
\beta(r)=\frac{1-{\dot t}}{{\dot r}}=\epsilon_r \frac{R(r^2f-ER^2-Lh)}{rf\sqrt{(ER^2+Lh)^2 + r^2f(\varepsilon R^2 - L^2)}}.\label{beta}
\end{align}
In the new coordinates $(T,r,\varphi)$, the metric (\ref{chargedBTZ}) is written as
\begin{align}
\D s^2=&-\frac{r^2}{R^2}\,f(\D T-\beta\D r)^2+f^{-1}\D r^2+R^2\biggl(\D\varphi+\gamma\D r+\frac{h}{R^2}(\D T-\beta\D r)\biggl)^2,\label{Doran-step}
\end{align}
where $\gamma$ and $\beta$ are given by Eqs.~(\ref{gamma}) and (\ref{beta}), respectively. Lastly, we impose a {\it key condition} ${g^{TT}=-1}$. This condition is satisfied with ${\varepsilon=-1}$, ${E=1}$, and ${L=0}$ for any set of $f(r)$, $h(r)$, and $R(r)$. Then, the transformations (\ref{T-def}) and (\ref{varphi-def}) and the metric (\ref{Doran-step}) coincide with Eqs.~(\ref{T-def2}), (\ref{varphi-def2}), and (\ref{BTZ-Doran}), respectively.

By the coordinate transformations (\ref{T-def2}) and (\ref{varphi-def2}) from Eqs.~(\ref{geo-BTZ1})--(\ref{geo-BTZ3}), we obtain geodesic equations for a timelike particle ({$\varepsilon=-1$}) with ${E=1}$ and ${L=0}$ in the Doran-type coordinates (\ref{BTZ-Doran}) as
\begin{align}
{\dot T}=&1,\qquad {\dot r}=\ \epsilon_r\frac{\sqrt{R^2 - r^2f}}{r},\qquad {\dot \varphi}=0.
\end{align}
Hence, ${T}$ is the proper time of the massive particle and ${\D r/\D T=\epsilon_r R/r}$ is satisfied on the Killing horizon ${f=0}$.
In particular, ${\D r/\D T=\epsilon_r=\pm 1}$ is satisfied for the solution with ${R=r}$ such as the rotating BTZ vacuum solution (\ref{rotatingBTZ-mj}).

\end{document}